\documentclass{article}

   \usepackage[final,nonatbib]{neurips_2024}

\usepackage[utf8]{inputenc} %
\usepackage[T1]{fontenc}    %
\usepackage{hyperref}       %
\usepackage{url}            %
\usepackage{booktabs}       %
\usepackage{amsfonts}       %
\usepackage{nicefrac}       %
\usepackage{microtype}      %
\usepackage{xcolor}         %
\usepackage{graphicx}
\usepackage{caption}
\usepackage{subcaption}
\usepackage{makecell}
\usepackage{multirow}
\usepackage{verbatim}

\newcommand{\wolf}[1]{{{\color{blue}{[\textbf{WG}: #1]}}}}
\newcommand{\todo}[1]{{{\color{red}{[\textbf{TODO}: #1]}}}}
\newcommand{\note}[1]{{{\color{magenta}{[#1]}}}} %

\newcommand{\hidecomment}[1]{} 
\newcommand{\mkclean}{
    \renewcommand{\note}{\hidecomment}
    \renewcommand{\todo}{\hidecomment}
    \renewcommand{\wolf}{\hidecomment}
}

\mkclean %

\newcommand{\introparagraph}[1]{\textbf{#1.}} %
\usepackage[capitalise,nameinlink]{cleveref} %

\title{Towards Agentic Schema Refinement}

\author{%
  Agapi Rissaki\thanks{Work done in part while the author was at RelationalAI.} \\
  Northeastern University\\
  \texttt{rissaki.a@northeastern.edu} \\
  \And
  Ilias Fountalis  \\
  RelationalAI \\
  \texttt{ilias.fountalis@relational.ai} \\
  \AND
  Nikolaos Vasiloglou \\
  RelationalAI \\
  \texttt{nik.vasiloglou@relational.ai} \\
  \And
  Wolfgang Gatterbauer \\
  Northeastern University\\
  \texttt{w.gatterbauer@northeastern.edu} \\
}

\begin{document}

\maketitle

\section{Introduction}
\label{introduction}

\introparagraph{The goal of a semantic layer}
Understanding the meaning of data is crucial for performing data analysis, yet for the users to gain insight into the content and structure of their database, a tedious data exploration process is often required \cite{inode, ma2023insightpilot}. A common industry practice taken on by specialists such as Knowledge Engineers is to explicitly construct an intermediate layer between the database and the user --- a \emph{semantic layer} --- abstracting away certain details of the database schema in favor of clearer data semantics \cite{semantic_scale, semantic_world}. 
In the era of Large Language Models (LLMs), industry practitioners and researchers attempt to circumvent this costly process using LLM-powered Natural Language Interfaces \cite{dong2023c3, gao2023text, li2023resdsql, pourreza2023din, rai2023improving, sun_sqlprompt_2023}. 
The promise of such Text-to-SQL solutions is to allow users without technical expertise to seamlessly interact with databases. For example, a new company employee could effectively issue queries in natural language 
without programming expertise or even explicit knowledge of the database structure, e.g., knowing the names of entities or properties, the exact location of data sources, etc. However, current Text-to-SQL solutions are far from perfect due to many factors, mainly related to ambiguous representations, such as data and query ambiguity \cite{huang2023data}, schema complexity (e.g., vast schema, wide tables, cryptic column names) \cite{floratou_nl2sql_2024}, among others. For these reasons, we argue that a semantic layer can be useful for human users and AI tools alike. 

\introparagraph{Contribution 1: Semantic layer as a set of distilled views}
We propose to define the semantic layer as a set of easy-to-interpret and reusable \emph{database views}. 
A view is a virtual table that is the result of a stored query, that can be used as any other table when writing subsequent queries \cite{cowbookEd3}. Database views are structured objects that can be constructed in-database, allowing for seamless integration with existing system components using direct communication in SQL. 
In particular, we explore the underlying database 
in order to discover views that represent entities together with their properties, as well as relationships between them. 
Thus, our semantic layer \emph{augments} a hard-to-interpret existing database schema with additional automatically generated and easy-to-interpret views, which distill the semantic knowledge into smaller bits. Note that this approach can also be viewed as a \emph{schema refinement process}, effectively reformulating the original tables into a larger set of more manageable ones, each with fewer columns, and clearly defined semantics.

\introparagraph{Contribution 2: Agentic programming for schema refinement}
Discovering such semantically meaningful database views is not a trivial task. If we had access to a vast log of past queries, we could attempt to mine a set of such views by targeting the queries or subqueries that appear frequently. Instead, we consider an even more challenging scenario where we only have access to the schema and the database. In this case, it seems impossible to recognize which parts of the data are semantically meaningful in an automated way; some form of external knowledge is necessary. Our solution uses LLMs to inject external knowledge into the process and to guide the view discovery.

We construct the semantic layer via an autonomous simulation process, leveraging LLM-powered multi-agent workflows \cite{guo_large_2024, li2024agenthospital}, i.e., \emph{agentic programming}. 
Such workflows incorporate feedback loops between agents, significantly enhancing their generative capabilities through iterative refinement \cite{mitra_agentinstruct_2024}. These systems can create complex simulations with minimal input, since agents autonomously generate \emph{both prompts and responses}, in contrast to a single-agent workflow where a human explicitly provides all prompts. Several tools have emerged to support agentic programming \cite{li2023camel, liu_agentlite_2024, wu_autogen_2023} by streamlining the development and coordination of multi-agent solutions. Further, agents can use external tools like APIs and code interpreters, which extend their functionality beyond LLMs, while advanced prompting techniques \cite{liu2023chain, shinn2023reflexion, wei2022chain, yao2023tree, yao2022react} further enhance agents' problem-solving abilities.

In our multi-agent approach, agents receive access to the database as well as minimal seed instructions and communicate to discover a set of views via a collaborative simulation. Specifically, we instruct the agents to use a systematic schema refinement mechanism (described in \cref{approach1}) and let them collaboratively apply the mechanism and to refine their results. The agents analyze the data, discover views, reflect on the quality of their proposals, and validate the view correctness using external tools. 

\introparagraph{Preliminary results}
We have applied our proposed method to several commercial databases and we show one case study in \cref{experiments}, illustrating structural and semantic properties of the distilled views. 

\section{Approach}
\label{approach}

We present our approach for semantic layer construction, via a simple systematic schema refinement mechanism for view discovery (\cref{approach1}), implemented as a multi-agent simulation (\cref{approach2}).

\subsection{Schema refinement mechanism} 
\label{approach1}

\begin{figure}
    \centering
    \includegraphics[width=\linewidth]{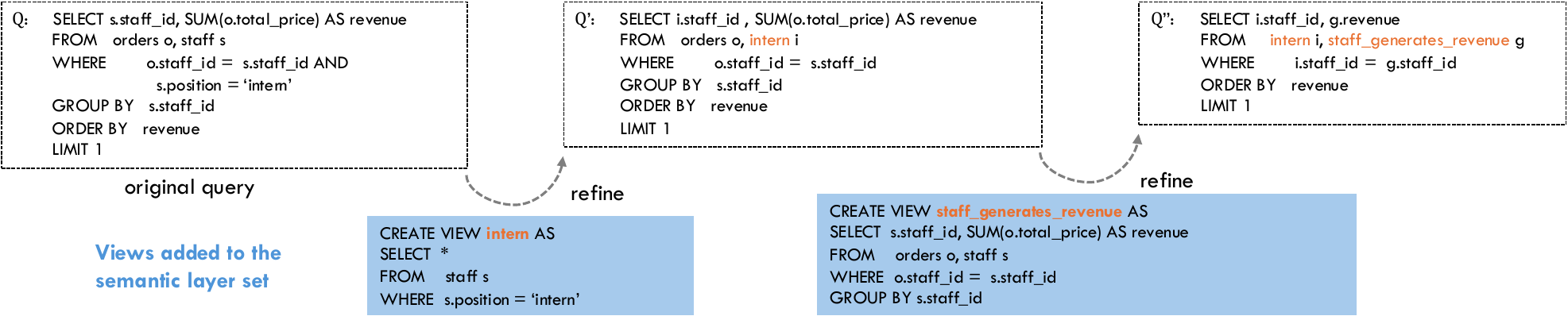}
    \caption{Example of the schema refinement mechanism on the schema \textit{orders( order\_id, staff\_id, total\_price, $\dots$), staff( staff\_id, position, $\dots$)}, where each order is handled by a staff member. By distilling the views \textit{intern} and \textit{staff\_generates\_revenue}, the query becomes successively more refined.}
    \label{fig:refinement}
\end{figure}

The main idea behind view discovery is to start with complex queries and then express them in a simpler way by decomposing them into modular components. These components are separately defined as database views that encapsulate distinct portions of the query logic, frequently representing key entities or relationships within the data. This \emph{query refinement} process effectively turns into \emph{schema refinement} by producing several reusable views as byproduct. 

We illustrate this process with a database about orders handled by staff members with the following simple schema: 
\textit{orders(order\_id, staff\_id, total\_price, $\dots$), 
staff(staff\_id, position, $\dots$)}. Each order is handled by a staff member and has a total price.
\Cref{fig:refinement} shows a query that involves joins, aggregations, and filtering conditions. 
The query selects staff members that are ``interns'', finds the orders handled by those staff members, calculates the total revenue produced by these orders per staff member, sorts staff members according to the total revenue generated and finally finds the intern that produced the highest revenue. Attempting to express this query in a simpler way, 
we could start by defining a view that corresponds to ``interns'' effectively omitting this selection from the original query. Note that the \emph{intern} can be interpreted as a distinct entity hidden in the database. We can further simplify this query by defining a view that precomputes the aggregate value ``revenue'' for each staff member. With this view we discovered a new property \emph{revenue} of the entity \emph{staff}. 

\subsection{Multi-agent framework} 
\label{approach2}

\begin{figure}
\begin{subfigure}[b]{0.48\linewidth}
    \centering
  \includegraphics
  [height=4cm]
  {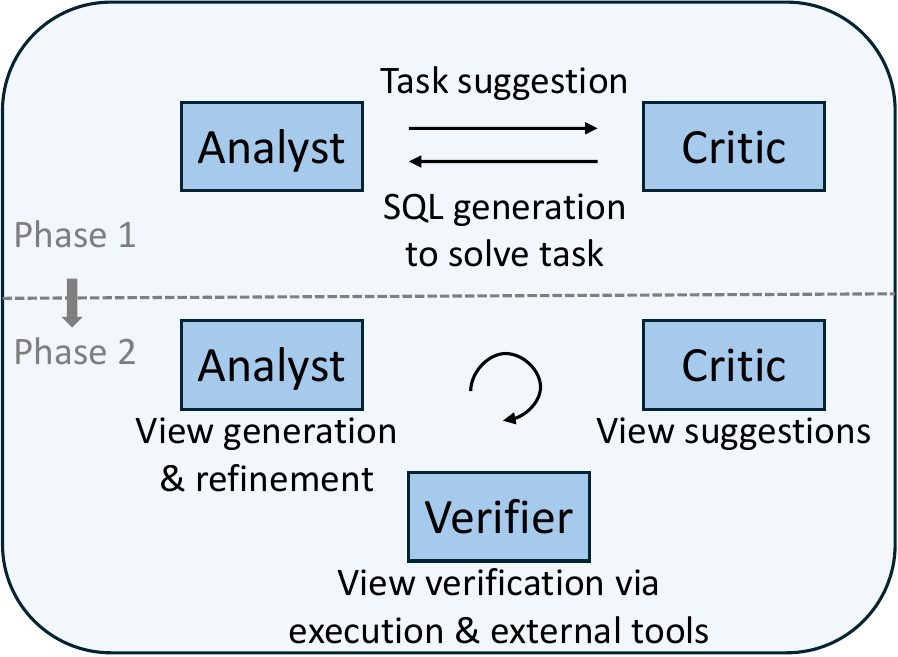}
  \caption{}
  \label{fig:schema_refinement_chat}
\end{subfigure}
\begin{subfigure}[b]{0.48\linewidth}
    \centering
  \includegraphics
  [height=4cm]
  {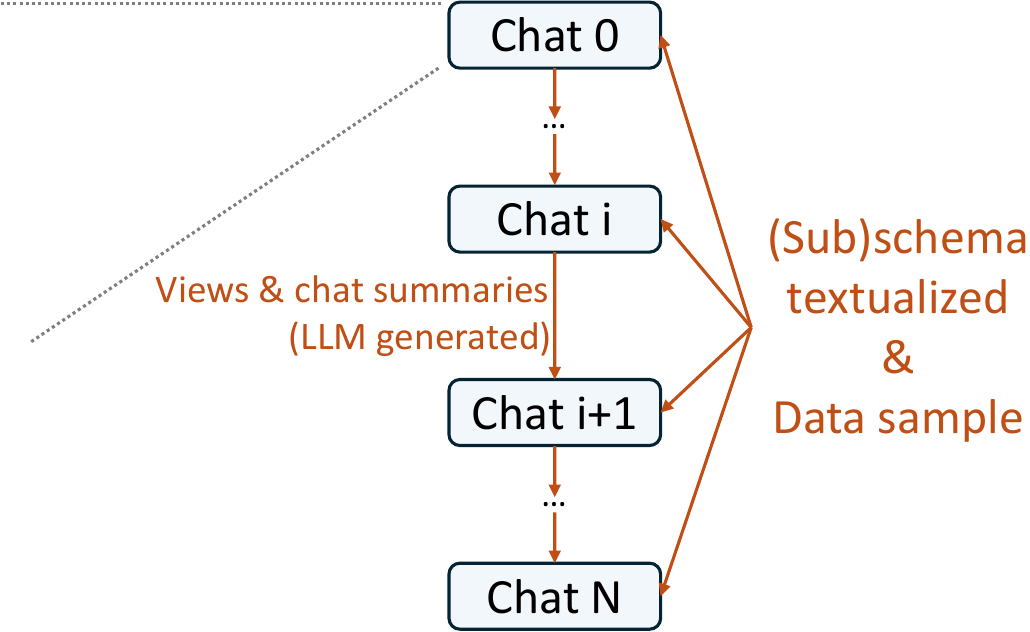}
   \caption{}
   \label{fig:sequential}
\end{subfigure}
\caption{A single chat session implementing the schema refinement mechanism (Figure (a)). A sequence of schema refinement chat sessions. Retaining memory from previous sessions helps promote view reusability (by using previously defined views in new tasks) and diversity (by avoiding defining the same views or working on similar tasks) (Figure (b)).
}
\end{figure}

In our multi-agent system three agents (Analyst, Critic, and Verifier) engage in iterative, free-form conversations and collaboratively implement the proposed schema refinement mechanism (\cref{fig:schema_refinement_chat}). More specifically, our agents have the following distinct roles:
\begin{itemize}
    
    \item The \emph{Analyst} implements analytics tasks, through the formulation of SQL queries and subsequently defines intermediate database views that refine these queries.

    \item The \emph{Critic} reviews these views to ensure that they are optimized with respect to query refinement, offering suggestions for improvement and proposing alternative formulations.
    
    \item The \emph{Verifier} validates the views by execution within a database engine, thereby ensuring their correctness, and may also use external tools for more sophisticated testing. 
\end{itemize}

Each simulation session receives as input a straightforward textual description of the database schema, supplemented by an optional small data sample. 
In most cases the schema is particularly large, thus the agents do not attempt to ingest it all at once. Instead, the process is performed on subsets of tables that join with each other, thereby encouraging the creation of views that integrate data from diverse sources. In particular, we sample \emph{connected components} of the \emph{schema graph}, i.e., the graph that represents tables as nodes and PK/FK pairs as edges. Further, we leverage a GraphRAG algorithm \cite{he_g-retriever_2024} to specialize the sampled subgraphs and thus guide the focus of each multi-agent session. 
Since the algorithm operates on graphs with features, we augment the schema graph with node and edge features that correspond to text embeddings 
of table and key descriptions, respectively. 

To achieve comprehensive coverage of the database, multiple sessions can be executed sequentially (\cref{fig:sequential}). Agents maintain a history of prior sessions to avoid redundancy and promote diversity in their exploration 
of the database schema. This ensures that the process not only scales effectively but also delivers a broad and diverse semantic understanding of the data.

\section{Experiments}
\label{experiments}

We report a preliminary study of our framework, quantifying the effects of our schema refinement on a realistic commercial database and showing how our results aid data exploration. We observe that our method provides a \emph{practical normalization} of the database into manageable components. 

\begin{figure}
\begin{subfigure}[b]{0.45\linewidth}
    \scriptsize
    \centering

    \begin{tabular}{l l || l l}
    \toprule
    \multicolumn{2}{c}{Original Schema} & \multicolumn{2}{c}{Semantic Layer} \\
     \# tables & 61 & \# views & 1146 \\ 
     \# median table width & 28 & \# median view width & 3 \\
     \# columns & 1770 & \# original columns & 1430 \\  
     \# relations & 27601 & \# relations & 22365 \\
    \bottomrule
    \end{tabular}
    \newline
    \newline
    \newline
    \caption{Comparison between original schema and semantic layer w.r.t. key statistics.}
    \label{tab:stats}
\end{subfigure}
\hfill
\begin{subfigure}[b]{0.45\linewidth}
    \centering
    \includegraphics[width=0.9\linewidth]{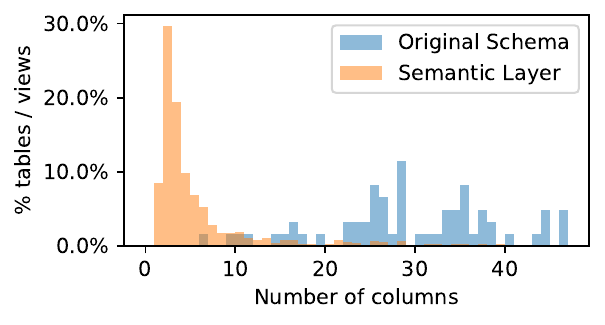}
    \caption{Table / view width distribution in original schema and semantic layer. }
    \label{fig:width}
\end{subfigure}
\caption{ Structural properties of the distilled views composing the semantic layer. We ignore the top 1\% views in terms of width.}
\label{fig:stats}
\end{figure}

\paragraph{Case Study.} We studied our approach on a realistic corporate database provided by Braze, a customer engagement platform, publicly available in Snowflake Marketplace \footnote{\tiny\url{https://app.snowflake.com/marketplace/listing/GZT0Z5I4XY0/braze-braze-user-event-demo-dataset}}. The database contains synthetic data and serves as a demo to Braze clients modelling user behavior and engagement with respect to events such as marketing campaigns. The provided schema consists of $61$ mostly wide tables  and $1770$ columns in total (\cref{tab:stats}). After a few hours of multi-agent simulation, we obtained a set of $1146$ views, covering $80.79\%$ of the columns in the original schema. Further, $54.84\%$ of the original relationships among columns (2 columns co-occurring in a table) are preserved in the views, while $7229$ \emph{newly} created pairwise relationships are obtained via views that combine data across different sources using join operators. 
In \cref{fig:width}, we see that most generated views have few columns (median width $3$, \cref{tab:stats}) in contrast to tables in the original schema which are often wide (median width $28$, \cref{tab:stats}), thus validating that the generated views serve as a refined version of the original schema with tables effectively decomposed into smaller components.

We then organize the generated views to better understand their semantics and provide easy user access. Specifically, we semantically group the views using embedding and clustering. Then we instruct an LLM-agent to discover entities, properties, relationships within groups and combine across groups, producing a comprehensive Entity-Relationship model. Note that each entity and relationship discovered is directly mapped to a set of database views, validated for correctness and usefulness via our multi-agent schema refinement framework. \Cref{tab:er_model} shows a snippet of our results. Starting from this collection, a new user can immediately gain insight about the semantic content of the database. Then they can locate exact data sources (tables / columns / derived attributes) by mapping each entity and relation to validated database views expressed in SQL within the database itself.

\begin{table}[b]
\caption{Entities and relationships mapped to discovered views.}
\label{tab:er_model}
\scriptsize
\centering
\begin{tabular}{| c | l | l | l |}
\hline
\makecell[l]{\textbf{Entity}  / \textbf{Relation}} & \textbf{Attributes} & \textbf{Views} & \textbf{Origin Tables} \\ 
\hline
\textbf{User} & \makecell[l]{language \\ location \\ ...} & \makecell[l]{Common\_User\_Attributes \\
User\_Language\_Distribution \\
User\_Demographics \\
...} 
& \makecell[l]{
USERS\_MESSAGES\_CONTENTCARD\_SEND \\
USERS\_BEHAVIORS\_UPGRADEDAPP \\
USERS\_BEHAVIORS\_UNINSTALL \\
USERS\_CAMPAIGNS\_CONVERSION \\
...
} \\
\hline
\textbf{Message} & \makecell[l]{content \\ type} & \makecell[l]{User\_SMS\_Behaviors \\ Spam\_Email\_Conversion\_Data \\ 
Non\_Spam\_Email\_Conversion\_Data \\ 
Email\_Conversion\_Data \\
In\_App\_Message\_Clicks} 
& \makecell[l]{USERS\_MESSAGES\_INAPPMESSAGE\_CLICK \\
USERS\_MESSAGES\_CONTENTCARD\_SEND \\
USERS\_MESSAGES\_EMAIL\_MARKASSPAM } \\
\hline
\textbf{Campaign} & \makecell[l]{conversion rates \\ metrics} & \makecell[l]{Push\_Notification\_Campaigns \\ 
Push\_Notification\_Conversion\_Rates} 
& \makecell[l]{USERS\_CAMPAIGNS\_CONVERSION \\
USERS\_MESSAGES\_PUSHNOTIFICATION\_SEND} \\
\hline
\hline
\textbf{User-Message} & - & \makecell[l]{User\_SMS\_Behaviors \\ User\_Message\_Variation\_Analysis} 
& \makecell[l]{USERS\_MESSAGES\_CONTENTCARD\_SEND \\
USERS\_MESSAGES\_EMAIL\_MARKASSPAM } \\
\hline
\textbf{User-Campaign} & - & \makecell[l]{Users\_Conversion\_Behavior, \\ Conversion\_Counts\_by\_Gender, \\
Campaign\_Conversion\_Counts} 
& \makecell[l]{USERS\_CAMPAIGNS\_CONVERSION} \\
\hline
\end{tabular}
\end{table}

\paragraph{Implementation.} Our multi-agent schema refinement solution is implemented using AutoGen \cite{wu_autogen_2023}, an open-source framework for agentic programming. LLM agents are powered by GPT-4 \cite{achiam2023gpt}. The code is available at \url{https://github.com/agapiR/agentic-semantic-layer}.

\section{Conclusion and Future Work}
\label{conclusion}

We proposed a view discovery approach to build a semantic layer within complex databases, implemented as a multi-agent simulation. In future work, we will use our semantic layer to boost text-to-SQL performance and we will integrate it into an interactive data exploration system. 

\bibliographystyle{plain}
\bibliography{references}

\newpage
\appendix

\section{Appendix}

\subsection{Case Study: CMS Data Feeds}
\label{cms}

In this section we present an additional case study, this time on data feeds from the Centers for Medicare and Medicaid Services \footnote{https://data.cms.gov/}. This is a very large diverse database with a star-like schema, where two tables \textit{Feeds} and \textit{Feed Files} are both in the center. 

\begin{table}
\centering
\begin{tabular}{l l || l l}
\toprule
\multicolumn{2}{c}{Original Schema} & \multicolumn{2}{c}{Semantic Layer} \\
 \# tables & 113 & \# views & 632 \\ 
 \# median table width & 24 & \# median view width & 4 \\
 \# max table width & 1130 & \# max view width & 31 \\
 \# columns & 6879 & \# original columns & 1298 \\  
 \# relations & 1121976 & \# relations & 7529 \\
\bottomrule
\end{tabular}
\caption{Comparison between original schema and semantic layer w.r.t. key statistics.}
\label{tab:cms_stats}
\end{table}

\begin{figure}
    \centering
    \includegraphics[width=0.7\linewidth]{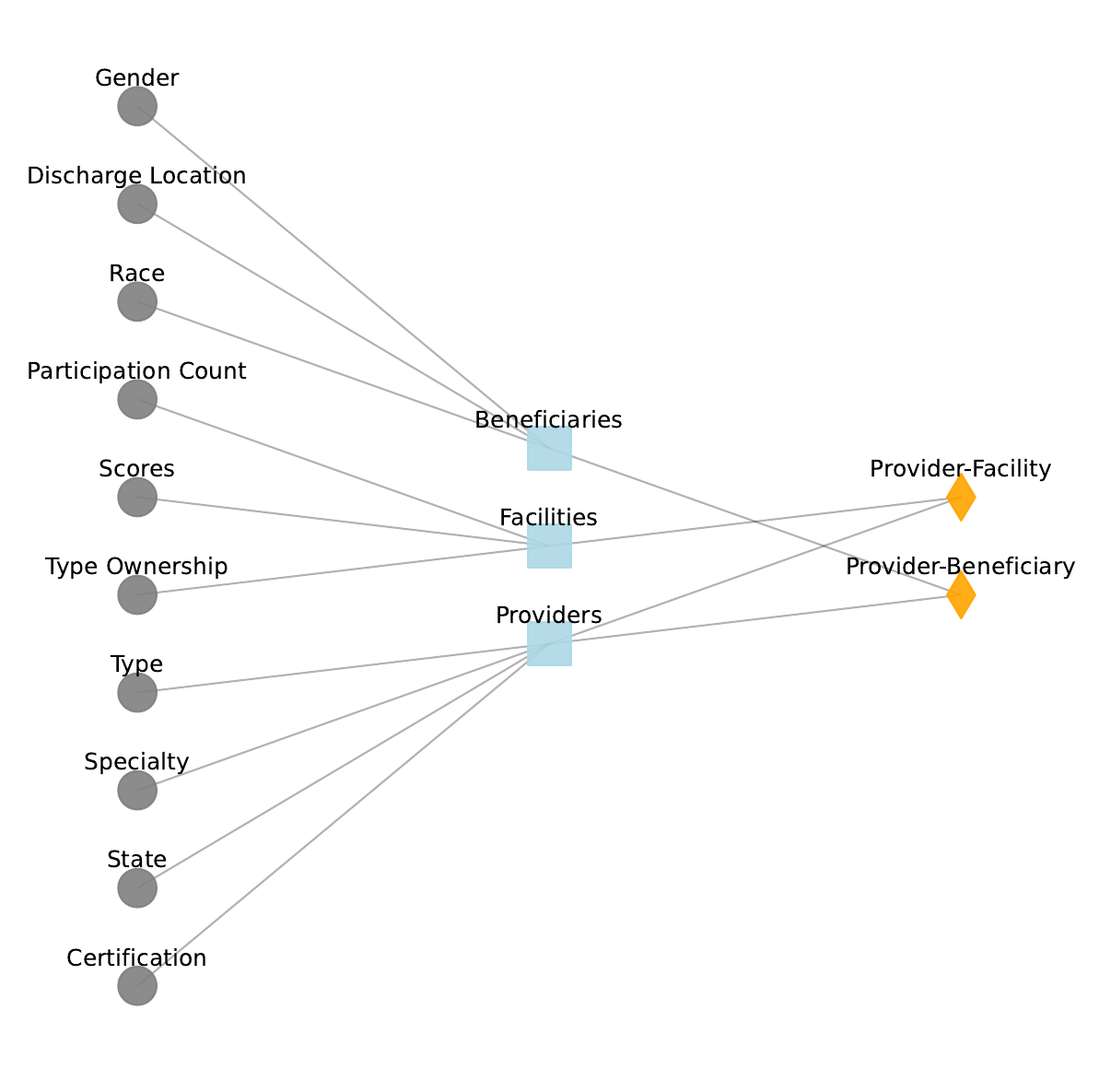}
    \caption{Diagram of entities (with their attributes) and relationships.}
    \label{fig:cms_er_diagram}
\end{figure}

\begin{table}[ht]
\caption{Entities and relationships mapped to discovered views.}
\label{tab:cms_er_model}
\scriptsize
\centering
\begin{tabular}{| c | l | l | l |}
\hline
\makecell[l]{\textbf{Entity}  / \textbf{Relation}} & \textbf{Attributes} & \textbf{Views} & \textbf{Origin Tables} \\ 
\hline
\textbf{Beneficiaries} & \makecell[l]{Gender \\ Discharge Location \\ Race } & \makecell[l]{beneficiaries\_gender\_distribution \\ beneficiary\_discharge\_location \\ beneficiary\_race\_distribution} 
& \makecell[l]{
MEDICARE\_COVID\_HOSPITALIZATION\_TRENDS, \\
MEDICARE\_POSTACUTE\_CARE\_AND\_HOSPICE, \\
\_BY\_GEOGRAPHY\_PROVIDER \\
MEDICARE\_INPATIENT\_HOSPITALS\_BY\_PROVIDER, \\
... 
} \\
\hline
\textbf{Providers} & \makecell[l]{Type \\ Specialty \\ State \\ Certification } & \makecell[l]{provider\_type\_state\_distribution \\ provider\_certification \\ providers\_by\_specialty} 
& \makecell[l]{
HOME\_INFUSION\_THERAPY\_PROVIDERS, \\
PROVIDER\_OF\_SERVICES\_FILE \\
\_CLINICAL\_LABORATORIES, \\
HOME\_HEALTH\_AGENCY\_ENROLLMENTS, \\
...
} \\
\hline
\textbf{Facilities} & \makecell[l]{Participation Count \\ Type Ownership \\ Scores} & \makecell[l]{facility\_participation\_count \\ facility\_type\_ownership \\ facility\_scores\_distribution} 
& \makecell[l]{
QUALITY\_PAYMENT\_PROGRAM\_EXPERIENCE,  \\
LONGTERM\_CARE\_FACILITY\_CHARACTERISTICS, \\
...
} \\
\hline
\hline
\textbf{Provider-Beneficiary} & - & \makecell[l]{beneficiaries\_by\_medical\_specialty} 
& \makecell[l]{
MEDICARE\_DURABLE\_MEDICAL\_EQUIPMENT \\
\_DEVICES\_SUPPLIES\_BY\_REFERRING\_PROVIDER \\
...
} \\
\hline
\textbf{Provider-Facility} & - & \makecell[l]{provider\_type\_home\_health\_agency} 
& \makecell[l]{
HOME\_HEALTH\_AGENCY\_ENROLLMENTS, \\
...
} \\
\hline
\end{tabular}
\end{table}

The schema consists of $113$ tables and $6879$ columns in total, for which we obtained a set of $632$ views. In \cref{tab:cms_stats}, we see that most generated views have few columns (median width $4$, maximum width $31$) in contrast to tables in the original schema which can be extremely wide (maximum width $1130$).

In this case, although the views cover $18.87\%$ of the columns in the original schema, abstracting away too detailed information, only $0.5\%$ of the original relationships among columns are preserved in the views. This is expected since 2 columns co-occurring in a table of extreme width do not necessarily reflect relations in the semantic level, thus the refinement process decomposes these kinds of tables preserving only key relations between columns and discarding the rest. Additionally, the views contain $1965$ \emph{newly} created pairwise relationships, associating columns from different origin tables.

Performing the same postprocessing as discussed in the main paper, we obtain an Entity-Relation model. In this case, many tables are extremely wide and table / column names are usually cryptic. Nonetheless, our method provides a comprehensive refined version of the database. For a small snippet see \Cref{tab:cms_er_model} and \Cref{fig:cms_er_diagram}.

\subsection{Sample multi-agent chat}
\label{chat}

\begin{footnotesize}
\begin{verbatim}
Critic, I have the following database schema.

/// SCHEMA WORDING OMITTED FOR BREVITY

Please suggest an analysis task for me to work on. 

--------------------------------------------------------------------------------
Critic (to chat_manager):

I suggest you analyze the behavior of users who have interacted with both email 
and push notification campaigns. Specifically, you can compare the engagement 
metrics (such as open rates, click-through rates, etc.) of these users across 
different campaigns and attributes. This analysis can provide insights into the 
effectiveness of your email and push notification strategies for different user 
segments. Let me know if you need more details or if you'd like to focus on a 
specific aspect of this analysis.

--------------------------------------------------------------------------------
Analyst (to chat_manager):

```sql
-- Count the number of users who have interacted with both email and push 
notification campaigns
WITH email_users AS (
    SELECT DISTINCT USER_ID
    FROM USERS_MESSAGES_EMAIL_SEND_VIEW
),
push_notification_users AS (
    SELECT DISTINCT USER_ID
    FROM USERS_MESSAGES_PUSHNOTIFICATION_IOSFOREGROUND_VIEW
)
SELECT COUNT(*) AS total_users
FROM email_users
JOIN push_notification_users 
ON email_users.USER_ID = push_notification_users.USER_ID;
```

```sql
-- Analyze the engagement metrics (open rates, click-through rates, etc.) 
of users who interacted with both email and push notification campaigns
WITH email_engagement AS (
    SELECT USER_ID, OPEN_RATE, CLICK_THROUGH_RATE
    FROM USERS_MESSAGES_EMAIL_SEND_VIEW
),
push_notification_engagement AS (
    SELECT USER_ID, OPEN_RATE, CLICK_THROUGH_RATE
    FROM USERS_MESSAGES_PUSHNOTIFICATION_IOSFOREGROUND_VIEW
)
SELECT AVG(email_engagement.OPEN_RATE) AS avg_email_open_rate,
AVG(email_engagement.CLICK_THROUGH_RATE) AS avg_email_click_through_rate,
AVG(push_notification_engagement.OPEN_RATE) AS avg_push_notification_open_rate,
AVG(push_notification_engagement.CLICK_THROUGH_RATE) AS avg_push_notification_click_through_rate
FROM email_engagement
JOIN push_notification_engagement 
ON email_engagement.USER_ID = push_notification_engagement.USER_ID;
```

--------------------------------------------------------------------------------
Critic (to chat_manager):

Your queries are a good start, but I recommend defining some views to simplify 
and enhance the analysis:

1. **View 1: users_interacted_with_email_push_campaigns**
   - This view will combine the distinct user IDs who 
   interacted with email and push notification campaigns.
   
   - It will be based on the common user interactions 
   between the two types of campaigns.

2. **View 2: email_engagement_metrics**
   - This view will extract relevant engagement metrics 
   (open rates, click-through rates, etc.) for email campaigns.
   - It will provide a cleaner and more focused dataset for
   email campaign analysis.

3. **View 3: push_notification_engagement_metrics**
   - This view will extract relevant engagement metrics for
   push notification campaigns.
   - It will allow for a clearer and more concise analysis 
   of push notification campaign engagement.

By creating these views, you can simplify your queries, make them more 
reusable, and ensure better readability. It's also a good practice to 
rename the columns in the views to make 
them more descriptive. Once you have these views in place, you can further 
analyze the engagement metrics or explore 
other aspects of user behavior. If you agree, please 
incorporate these views into your analysis. Let me know if you need any 
further clarification or assistance.

--------------------------------------------------------------------------------
Analyst (to chat_manager):

```sql
-- Creating View 1: users_interacted_with_email_push_campaigns
CREATE VIEW users_interacted_with_email_push_campaigns AS
SELECT USER_ID
FROM USERS_MESSAGES_EMAIL_SEND_VIEW
WHERE USER_ID IN (
    SELECT USER_ID
    FROM USERS_MESSAGES_PUSHNOTIFICATION_IOSFOREGROUND_VIEW
);

-- Creating View 2: email_engagement_metrics
CREATE VIEW email_engagement_metrics AS
SELECT USER_ID, OPEN_RATE AS email_open_rate, 
       CLICK_THROUGH_RATE AS email_click_through_rate
FROM USERS_MESSAGES_EMAIL_SEND_VIEW;

-- Creating View 3: push_notification_engagement_metrics
CREATE VIEW push_notification_engagement_metrics AS
SELECT USER_ID, OPEN_RATE AS push_notification_open_rate, 
       CLICK_THROUGH_RATE AS push_notification_click_through_rate
FROM USERS_MESSAGES_PUSHNOTIFICATION_IOSFOREGROUND_VIEW;
```

```sql
-- Analyzing the combined engagement metrics of users who 
-- interacted with both email and push notification campaigns
SELECT AVG(email.email_open_rate) AS avg_email_open_rate,
AVG(email.email_click_through_rate) AS avg_email_click_through_rate,
AVG(push.push_notification_open_rate) AS avg_push_notification_open_rate,
AVG(push.push_notification_click_through_rate) AS avg_push_notification_click_through_rate
FROM email_engagement_metrics email
JOIN push_notification_engagement_metrics push 
ON email.USER_ID = push.USER_ID;
```  

--------------------------------------------------------------------------------
Critic (to chat_manager):

Your approach to creating views for users who interacted with both email and 
push notification campaigns, as well as extracting engagement metrics for analysis, 
is well-structured and effective. The views you created will not only simplify 
your queries but also make them more reusable and maintainable. 

I appreciate how you've renamed the columns in the views to make them more 
descriptive and readable. This will help improve the overall quality of 
your analysis and make it easier for users to understand the data being presented.

Your final analysis query combining the engagement metrics of users who 
interacted with both types of campaigns looks solid. It provides a clear 
comparison between email and push notification engagement metrics, allowing 
for insights into user behavior across different communication channels.

Keep up the good work!

--------------------------------------------------------------------------------
Analyst (to chat_manager):

Thank you for your feedback and guidance. I'm glad to hear that the views and 
analysis approach meet your expectations. 

--------------------------------------------------------------------------------
Critic (to chat_manager):

Great job. Goodbye.

--------------------------------------------------------------------------------
Verifier (to chat_manager):

*** Suggested tool call (call_4dz4c1irNRJiOdrijI2qkmB7): materialize_view_tool ***
Arguments: 
{"view_definitions":["CREATE VIEW users_interacted_with_email_push_campaigns 
AS SELECT USER_ID FROM USERS_MESSAGES_EMAIL_SEND_VIEW WHERE USER_ID IN 
(SELECT USER_ID FROM USERS_MESSAGES_PUSHNOTIFICATION_IOSFOREGROUND_VIEW)", 
"CREATE VIEW email_engagement_metrics AS SELECT USER_ID, OPEN_RATE 
AS email_open_rate, CLICK_THROUGH_RATE AS email_click_through_rate 
FROM USERS_MESSAGES_EMAIL_SEND_VIEW", 
"CREATE VIEW push_notification_engagement_metrics 
AS SELECT USER_ID, OPEN_RATE AS push_notification_open_rate, 
CLICK_THROUGH_RATE AS push_notification_click_through_rate 
FROM USERS_MESSAGES_PUSHNOTIFICATION_IOSFOREGROUND_VIEW;"]}
**********************************************************************************

--------------------------------------------------------------------------------

>>>>>>>> EXECUTING FUNCTION materialize_view_tool...

Verifier (to chat_manager):

***** Response from calling tool (call_4dz4c1irNRJiOdrijI2qkmB7) *****
["View successfully defined.", 
"View successfully defined.", 
"View successfully defined."]
**********************************************************************

--------------------------------------------------------------------------------
\end{verbatim}
\end{footnotesize}

\end{document}